\documentclass[runningheads]{llncs}
\usepackage[T1]{fontenc}
\usepackage{graphicx}

\usepackage{soul}
\usepackage{todonotes}
\usepackage{xcolor}
\usepackage{amsmath}
\usepackage{hyperref}
\usepackage{comment}
\usepackage{listings}
\usepackage{tabularx}

\usepackage{subcaption} 

\usepackage{paralist}
\usepackage{enumitem}
\usepackage{amssymb}

\usepackage{longtable}
\usepackage{booktabs}

\newcommand{\ignore}[1]{}

\begin{document}
\title{Time and Relations into Focus:\\ Ontological Foundations of\\ Object-Centric Event Data
}

\titlerunning{Ontological Foundations of Object-Centric Event Data}

\institute{Anonymous}
    
\author{Hosna Hooshyar\inst{1} \and
Mattia Fumagalli\inst{1} \and
Marco Montali\inst{1} \and Giancarlo Guizzardi \inst{2}}
\authorrunning{H.~Hooshyar et al.}
\institute{Free University of Bozen-Bolzano, Italy\\
\and
Semantics, Cybersecurity {\usefont{OT1}{cmtt}{m}{it} \&} Services (SCS), University of Twente, The Netherlands
}
%
\maketitle              
\begin{abstract}

Object-centric process mining is a new branch of process mining where events are associated with multiple objects, and where object-to-object interactions are essential to understand the process dynamics. Traditional event data models, also called \textit{case-centric}, are unable to cope with the complexity introduced by these more refined relationships. Several models have been made to move from case-centric to \textit{Object-Centric Event Data (OCED)}, trying to retain simplicity as much as possible. Still, these suffer from inherent ambiguities, and lack a comprehensive support of essential dimensions related to time and (dynamic) relations. In this work, we propose to fill this gap by leveraging a well-founded ontology of events and bringing ontological foundations to OCED, with a three-step approach. First, we start from key open issues reported in the literature regarding current OCED metamodels, and witness their \textit{ambiguity} and \textit{expressiveness limitations} on illustrative and representative examples proposed therein. 
%
%
Second, we consider the \textit{OCED Core Model}, currently proposed as the basis for defining a new standard for object-centric event data, and we enhance it by grounding it on a lightweight version of \textit{UFO-B} called gUFO, a well-known foundational ontology tailored to the representation of objects, events, time, and their (dynamic) relations. This results in a new metamodel, which we call gOCED. The third contribution then shows how gOCED at once covers the features of existing metamodels preserving their simplicity, and extends them with the essential features needed to overcome the ambiguity and expressiveness issues reported in the literature.  

\end{abstract}

\section{Introduction}
\label{introduction}

Process mining aims to provide fact-based, actionable insights that support the 
improvement of organizational processes~\cite{Dealing_with_divergence_convergence}. Its methods rely on event logs, recording events extracted from information systems that provide a digital footprint of the actual execution of business/work processes~\cite{OC_PM}. Traditional process mining techniques assume that each event refers to exactly one object (called \textit{case}), such as 
a parcel. Based on this assumption, events relating to the same case are grouped together, yielding a trace of events that describes the evolution of that case through the process, called a process instance. As a result, every event belongs to exactly one case, and each case is analyzed in isolation~\cite{Uncovering_OCED}.

This strong assumption has been subjected to extensive criticism lately. In fact, real-world processes often inherently exhibit a much more complex behavior: events may refer to multiple, inter-related objects that co-evolve by participating in one or multiple concurrently running processes~\cite{Framework_Extracting_Encoding}. To address this complexity, \textit{Object-Centric Process Mining} is emerging as a field of study where event data and corresponding process mining tasks are lifted to this richer setting. This calls for (event) data modelling capabilities, and corresponding analysis techniques, able to more accurately capture the interactions among multiple entities, for instance by representing one-to-many and many-to-many \textit{Event-to-Object (E2O)}  and \textit{Object-to-Object (O2O)} relationships ~\cite{OCPM_Introduction}.

Despite recent progress, the literature indicates that a widely accepted standard model for representing \textit{Object-Centric Event Data (OCED)} has yet to emerge. Notable efforts in this direction include \textit{Object-Centric Event Logs (OCELs)}~\cite{OCELV1,OCEL2} (lifting event logs to the object-centric case),  \textit{Event Knowledge Graphs (EKGs)}~\cite{Multiple_Behavioral_Dimensions_EKG,Implementing_OCED_EKG} (representing OCEDs in a graph-structured way), and the \textit{OCED Core Model}~\cite{TowardsStandardOCED}, which stemmed from OCELs and EKGs to provide a common minimal basis for such a new standard.
While addressing essential characteristics of OCEDs, all such approaches remain \textit{ambiguous} and lack sufficient \textit{expressivity}, since: (1) their semantics are often under-specified, in turn forcing ad-hoc decisions when defining mining and analysis tasks, and (2) they lack first-class constructs to deal with relevant scenarios. To substantiate this claim, consider a basic example on student supervision, inspired by~\cite{TowardsStandardOCED}.

\begin{example}
\label{ex:supervision}
Consider three objects: a student, $S$, and two professors, $M$ and $D$. Two events, $e1$ and $e2$, respectively occur at timestamps $Time_1$ and $Time_2$ (with $Time_1< Time_2$), and record that $S$ is assigned to $M$, and that $S$ is assigned to $D$. The following event table represents this log following the OCEL format:

\begin{center}
    \begin{tabularx}
    {\textwidth}{|*{4}{>{\raggedright\arraybackslash}X}|}
    \multicolumn{1}{l}{\textbf{Event ID}}
    &
    \textbf{Event Type}
    &
    \textbf{Timestamp}
    &
    \multicolumn{1}{l}{\textbf{Related Objects}}
    \\
    \hline
    $e1$ & assign student & $Time_1$ & $M$,$S$
    \\
    \hline
    $e2$ & assign student & $Time_2$ &
    $D$,$S$
    \\
    \hline
    \end{tabularx}
    \end{center}

This table is encoded into a corresponding EKG using the methodology in \cite{Multiple_Behavioral_Dimensions_EKG}:\footnote{For simplicity, we momentarily neglect here derived event-to-event relationships.} \begin{inparaenum}[\itshape (i)]
\item the two events become two distinct event nodes (decorated with their timestamps);
\item the three objects become three distinct object nodes;
\item whenever an object occurs with an event, the corresponding nodes are connected with a generic E2O relationship;
\item whenever two objects participate in the same event, the corresponding nodes are connected with a generic O2O relationship
\end{inparaenum}
The resulting graph is shown in~\autoref{fig:supervisionScenario}, in a way that conforms to the \textit{OCED Core Model}.

\begin{figure}[t]
    \centering
    \includegraphics[width=0.85\linewidth]{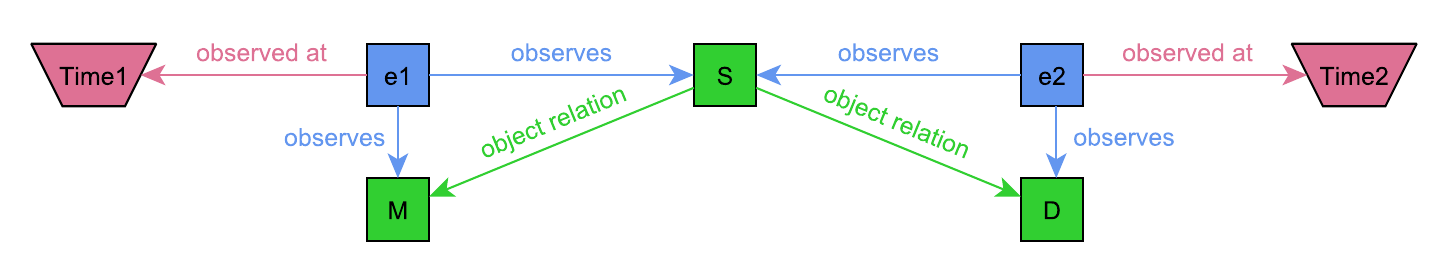}
    \caption{Graph representation of Example 1, according to the OCED Core Model.}
    \label{fig:supervisionScenario}
\end{figure}


Regardless of which of these representations is selected, what these two events do with respect to the related objects is inherently \emph{ambiguous}, as it is not clear how the supervision relationship is manipulated. First, by looking into the natural language description of the event types, one may only reasonably ``guess''  that $e1$ establishes that $M$ supervises $S$. Second, even under this interpretation, it is not clear whether $e2$ is \emph{adding} $D$ as supervisor, or \emph{replacing} $M$.

This problem also highlights the need for further \emph{expressiveness}. In fact, none of the aforementioned approaches provides abstractions to represent the impact of events on objects and relationships, in particular when it comes to time-varying properties (with the exception of time-varying attributes supported by OCEL V2) \cite{OCEL2}.
In addition, key aspects (further elaborated in Section \ref{OCEDchallenges}) cannot be represented in this approach. For example, it is not possible to single out object roles, 
nor the nature of relationships, e.g., to clarify whether $M$ and $D$ are in two distinct supervisory relationships with $S$, or instead co-participate to the same supervision. The only patch is to attempt capturing these aspects through reification, that is, by introducing additional, dedicated objects. However, this is only shuffling these issues around, since it is neither possible to classify and ascribe intrinsic properties to such objects, nor to differentiate the nature of an object, e.g., whether it is an independent object (e.g., a person), the state of an object (e.g., a canceled order), an intrinsic aspect of an object that may change (e.g., the language proficiency of an employee), or a relational entity that is existentially depends on multiple objects (e.g., a contract).
\end{example}

Two issues emerge when handling ambiguity and lack of expressiveness as shown in Example~\ref{ex:supervision}. First, further ad-hoc assumptions are made to disambiguate event data; for example, several techniques rely on the implicit assumption that O2O relations do not change over time \cite{SWGM25}. Second, depending on the requirements of the target applications, bottom-up extensions are proposed to handle the missing features. For example, the authors of \cite{Wei2025DirigoAM} extended OCEL V2 to capture the temporal aspect of dynamic object relationships, and distinguished between static and dynamic object attributes. Likewise, \cite{Khayatbashi2025AIEnhancedBP} defined a validity period for O2O relations within OCEL V2 for their case study. 

The aim of this work is to address these key concerns in a more systematic way, by grounding OCED into a suitable foundational ontology. To do so, we proceed in three steps. 
First, we start from the key open challenges raised in the OCED standard proposal~\cite{TowardsStandardOCED}  regarding current OCED metamodels. 
Second, we enhance the \textit{OCED Core Model} by grounding it in UFO-B, a suitable subset of the well-established \textit{Unified Foundational Ontology (UFO)} (the upcoming ISO/IEC CD 21838-5) ~\cite{guizzardi2022ufo,botti2019representing,gufo_github} that covers the essential dimensions we need: objects, events, time, and (time-evolving) attributes and relationships, with the possibility of describing their intrinsic properties in a 
explicit way. 
 More specifically, we rely on a lightweight version of UFO-B, called \emph{gUFO} \cite{gufo_github}, as it suffices for the features we need, and in turn provides a good trade-off between simplicity and expressiveness. We call the resulting metamodel \emph{gOCED}.
As our third and last contribution, we show that gOCED at once covers the features of existing metamodels preserving their simplicity, and extends them with the essential features needed to overcome the ambiguity and expressiveness issues reported in the literature. Importantly, our approach is non-intrusive. In fact, existing logs can be directly represented in gOCED without additional effort, and with the advantage that implicit assumptions are now made explicit. Then, additional features are systematically provided in gOCED thanks to its grounding in gUFO, thus avoiding to resort to ad-hoc, representational tricks.

All in all, gOCED establishes an unexplored bridge between two neighboring, often disconnected lines. As for process mining, it provides a well-founded way to represent rich object-centric event data. At the same time, it grounds a foundational ontology into a concrete and relevant application domain.

The paper is structured as follows. \autoref{OCEDchallenges} introduces OCED and its main open challenges. \autoref{related} discusses related work. \autoref{UFOB} recalls UFO-B and illustrates the subset we employ in the context of this paper. \autoref{mapping} describes our grounding of the OCED Core Model in gUFO. \autoref{eval} provides a conceptual assessment of the proposal, showing how the resulting metamodel addresses the identified representation challenges. \autoref{empiricalEval} presents an empirical evaluation based on the instantiation of the metamodel on a case-study dataset. Finally, \autoref{conclusion} concludes the paper and discusses future perspectives.

\section{The OCED Effort and Main Open Challenges} 
\label{OCEDchallenges}

\begin{figure}[t]
    \centering
    \includegraphics[width=0.8\textwidth]{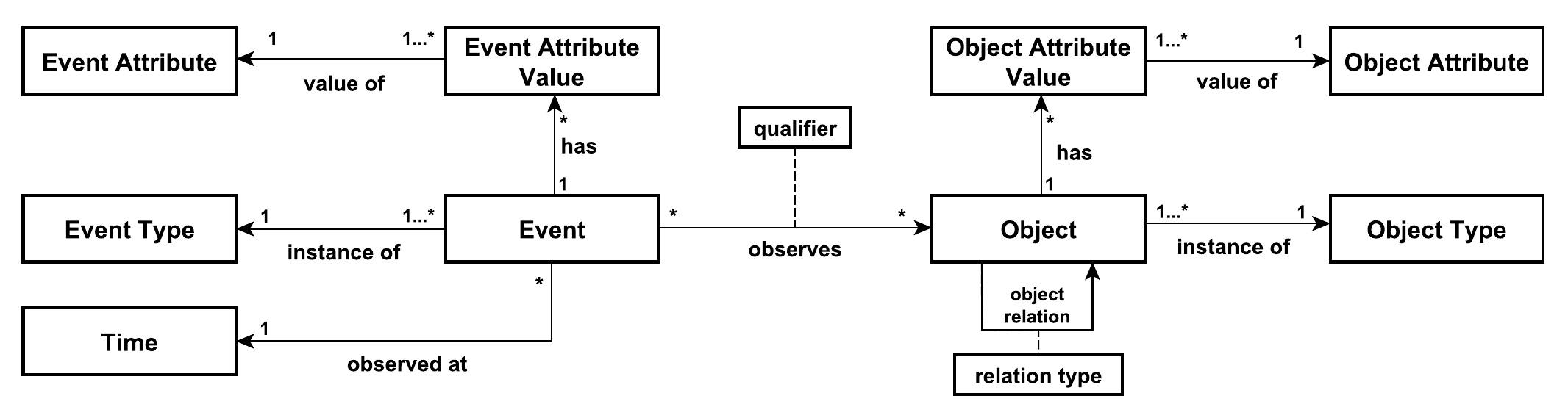}
    \caption{OCED Core Model proposed by Fahland et al.~\cite{TowardsStandardOCED}}
    \label{fig:coreModel}
    \vspace{-1em}
\end{figure}

The OCED community effort, summarised in~\cite{TowardsStandardOCED}, provides the basis for a new standard for object-centric event data for process mining. \autoref{fig:coreModel} illustrates the core OCED metamodel, which covers all the essential features of some working proposals for object-centric event data, such as OCEL and EKG\footnote{The only missing feature being that of time-varying attributes from OCEL V2}. We refer to this metamodel as the \textit{``Core Model''} throughout this document. Essentially, the Core Model states that each event can be associated to various objects through an E2O relationship, and that objects can be associated to each other through O2O relationships. Objects and events are typed, and can be decorated with attributes. Each event also comes with a specific attribute denoting its punctual timestamp. Finally, E2O and O2O relationships can be qualified (to denote their types or other key details).

Summarizing the state of the art and the outcome of community gatherings,  \cite{TowardsStandardOCED} singles out several key limitations of the Core Model, which we recall next.

\noindent \textbf{C1}. \textit{No relations between events}. The OCED Core Model lacks explicit \textit{Event-to-Event (E2E)} relationships. Events can be complex and may require to be interconnected, for example through part-whole relationships, causality, partial orders, as well as (object-mediated) directly-follows relations \cite{Multiple_Behavioral_Dimensions_EKG}. For example, an event of \textit{buying} may trigger a subsequent event of \textit{shipping}. Likewise, the existence of one event may depend on another; for instance, a \textit{Paying} event cannot occur unless a preceding event has issued a receipt. 

\noindent \textbf{C2}.
\textit{Event atomicity}. 
According to the Core Model, events are punctual and associated to a single time point. However, in real-world scenarios, events can be complex and span a duration. For instance, the event of \textit{shipping orders} may comprise several sub-events, such as \textit{packing items into boxes}, and \textit{loading boxes onto trucks}. Clearly, such a composite event may take anywhere from several hours to multiple days to complete. 
This limitation is typically addressed by mapping a durative event into two punctual events denoting start and completion \cite{xes,OCEL2}.

\noindent \textbf{C3}. \textit{Representing objects over time}. The characteristics of objects may change because of certain interactions between objects or because certain events affect the object. 
For example, an Order's status may transition from \textit{Shipped} to \textit{Delivered}. The Core Model does not specify how dynamic attribute values should be handled when the object itself remains unchanged, nor does it define how different observations of the same object are related to one another.

\noindent \textbf{C4}. \textit{Binary O2O}. In the Core Model, only binary object relationships are supported. However, there are situations where higher-arity relationships may be required. For instance, a customer may buy multiple products within the same order.
Reifying the tuple formed by the objects participating in O2O is still not enough to solve the issue, because the conventional notation for associations collapses two distinct kinds of multiplicity constraints \cite{Guizzardi2008WhatsIA}.




\noindent \textbf{C5}. \textit{The identification of O2O}. In this metamodel, O2O does not possess an identifier and is implicitly and uniquely identified by the pair of objects involved, while the relationship type indicates the nature of the connection. Under a strict interpretation, this implies that only one type of relationship can exist between any two objects. 
Some metamodels such as OCEL V2 do not adhere to this restriction and identify a relation using a triple consisting of two involved objects, and relation type \cite{OCEL2}. However, this alone does not resolve the issue, as multiple relationships of the same type can exist between the same objects. A person can have more than one employment contract with the same institution, one for normal working days and one for night hours and weekends.

\noindent \textbf{C6}. \textit{O2O Ambiguity}. The OCED Core Model only represents the existence of O2O, without capturing its attributes or dynamic changes. For instance, consider an order comprising multiple items. The relationship between the order and a specific item ceases to exist when it is removed from the order. 


The issues outlined above illustrate the potential ambiguity and limited expressiveness of the OCED Core Model. 
It is worth noting that the Core Model (as well as related efforts such as OCEL and EKG) can, in principle, approximate the aspects 
mentioned above, but at the cost of introducing extra modeling conventions and ad-hoc techniques. 

In this context, leveraging an ontologically well-founded conceptualization plays a key role, guiding modelers toward standardized ways of understanding and representing the phenomenon at hand, promoting interoperability and re-usability, and helping to avoid the burden of unintended modeling choices and ad hoc worka\-rounds, in addition to supporting a suitable cognitive understanding of the relations between objects, events and their properties aligned to the semantics of these entities as they occur in the world. In the following sections, we show how grounding the model in a well-founded ontology can address the mentioned challenges and limitations of the Core Model, while preserving its core functionalities and key features.

\section{Related Work}
\label{related}

Given the importance of an OCED metamodel, several efforts have been made to propose suitable representations, without leveraging a foundational ontology.


\cite{ocelV1_web,OCELV1} introduced 
OCEL V1, dealing with events, objects, their attributes, and many-to-many E2O. However, it does not account for E2E, O2O, dynamic attribute values, or any characteristics of E2O.
OCEL V1 was extended in \cite{DOCEL} into \textit{Data-aware OCEL (DOCEL)}, supporting attributes with multiple simultaneous values, dynamic attribute values, and unambiguous linking of attributes to either objects or events. Dynamic attributes are linked to  objects and the events that change their  values. The OCEL V1 framework was questioned in
\cite{Unraveling_Fabric}, forming the basis for OCEL V2 \cite{OCEL2}. In this model, events continue to be atomic, and E2E relations are still not supported. 
To represent activity instances spanning a duration, the author suggests using separate start and end events or adding a duration attribute. 
E2O and O2O relations can now include a qualifier, 
and O2O are still modelled as static. 
Attributes are linked to types rather than directly to objects or events, and object attributes can be dynamic and have timestamps. The latter feature supports time-varying attributes without forcing to explicitly anchor them to events.

Event-Knowledge Graphs (EKGs) were initially introduced in \cite{Multiple_Behavioral_Dimensions_EKG} 
to provide a graph-structured representation for OCEDs. An EKG supports multiple directly-follows relations across events associated with different entities, covering (static) O2O, E2E, and E2O relations. In \cite{TowardsStandardOCED}, EKGs and property graph schemas were used to provide a concrete storage for data obeying the OCED full metamodel from \cite{Implementing_OCED_EKG}.  
The implementation supports (static) O2O and additionally incorporates E2E to express \textit{directly-follows} dependencies between events.

Finally, iDOCEM is an Integrated Data and Object-Centric metamodel which aligns the concepts of DOCEL with an artifact\--centric modeling approach named ME\-RODE3 \cite{Goossens2023AligningOE,Verbruggen2024iDOCEMDA}. 
The iDOCEM is organized into three sections: (1)objects and their attributes; (2)activities and events; and (3)methods and parameters that link events to objects. iDOCEM models O2O and E2O relationships, but does not include E2E relations at the model level. 


\section{Leveraging a Well-founded Ontology of Events}
\label{UFOB}

To overcome the mentioned limitations, we adopt an established theory specifically designed to model events, objects, their properties, and relationships. This theory is a subset of the \textit{Unified Foundational Ontology (UFO)} (the upcoming ISO/IEC CD 21838-5) \cite{guizzardi2022ufo} named \textit{UFO-B} \cite{almeida2019events,guizzardi2013towards,guizzardi2016ontological}. UFO is an axiomatic theory that aggregates results from philosophy, cognitive science and linguistics, and has been tested in a multitude of real-world scenarios over the past decades (see \cite{guizzardi2022ufo}). For our purposes, we focus on a specific subset of UFO-B, which has been implemented in \textit{gUFO} \cite{gufo_github}, which is a lightweight, OWL-based ontology engineered to facilitate the application of UFO-B in diverse domains, including \textit{Semantic Web technologies} such as \textit{OWL 2 DL}\footnote{https://www.w3.org/TR/owl2-overview/}. 

gUFO reports the main UFO-B foundational distinction between \texttt{Endurant}s and \texttt{Event}s. Endurants are entities that are wholly present at a given moment in time, and which can qualitatively change while maintaining their identity (e.g., you and me, the United Nations, but also the marriage between John and Mary, as well as John's ability to speak Spanish). Events, also referred to as \texttt{Perdurants}, contrast with endurants in that they unfold over time and accumulate temporal parts, e.g., \textit{shopping processes}.

\begin{figure}[t]
    \centering
    \includegraphics[width=0.9\textwidth]{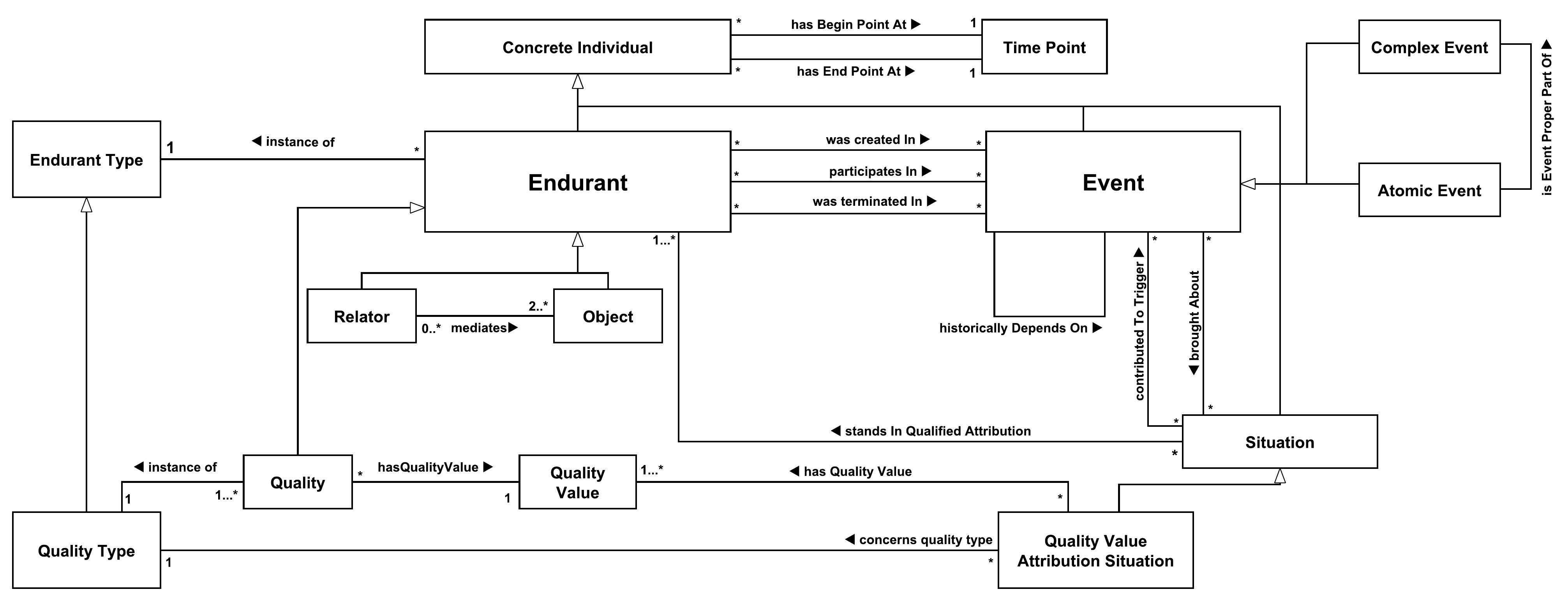}
    \caption{A summary of UFO-B and its relations as from~\cite{almeida2019events} and \cite{guizzardi2013towards}.}
    \label{fig:UFOB}
\end{figure}

Accordingly, gUFO retains six major aspects related to the ontological nature of events modeled in UFO-B \cite{guizzardi2013towards,almeida2019events}. As a first aspect, it accounts for the mereological structure of events. Events can be structured into part-whole hierarchies 
based on their internal composition, following what is termed an extensional mereology, namely, a partial ordering relation, plus the so-called weak supplementation axiom (i.e., if $e_2$ is part of $e_1$, then there is at least another $e_3$ that is disjoint from $e_2$ and that is also part of $e_1$) and the extensionally axiom (two events are the same iff they have the same parts). This enables the differentiation between atomic and complex events, as shown in the \texttt{Event}, \texttt{Complex Event}, and \texttt{Atomic Event} branches of \autoref{fig:UFOB}. 

Secondly, the model covers the existential dependence of events on endurants and formalizes the concept of \texttt{participation}. The maximal part of an event that depends exclusively on a specific endurant is a participation of that endurant in the event. In this view, an event can be partitioned into a set of mutually disjoint and exhaustive participations. For example, in a customer purchasing process, one can distinguish the participation of customer, dealer, and item. 

Third, the model covers the temporal nature of events. Events unfold through time by accumulating temporal parts, and their defining features are primarily temporal. Hence, all Allen interval relations \cite{allen1983maintaining} can be used to relate them. 

Fourth, events are manifestations of \texttt{Aspect}s that inhere in \texttt{Object}s (i.e., object-like entities that are existentially dependent on other objects). Of particular interest are intrinsic aspects such as \texttt{Qualities} (e.g., the height of a person), which can be triggered by a particular situation, resulting in the occurrence of an event. This interrelation is represented via the \texttt{Event} and \texttt{Situation} classes. 


As a fifth aspect, the model treats events as transitions from one \texttt{Situation} (i.e., a part of the world that can be understood as a whole; a spatiotemporally localized state of affairs) to another. Events are triggered by specific situations that activate the relevant qualities, and they result in new situations brought about by their occurrence (see \texttt{Event} and \texttt{Situation} in \autoref{fig:UFOB}). 

Finally, UFO-B introduces \textit{event types}. Events are not only individual occurrences but  also denote instances of broader event classes. This categorization supports abstraction, classification, and reasoning over types of events.

\section{Grounding OCED into gUFO}
\label{mapping}

This section outlines the construction of the OCED ontological metamodel grounded in gUFO, hereafter also \textit{gOCED}. To achieve this objective, we survey the core concepts of OCED, singling out their counterparts in gUFO.

\vspace{0.2em}
\noindent$\blacktriangleright$ \texttt{Object}.
The Core Model defines an object as an entity that represents either a tangible or an abstract element \cite{TowardsStandardOCED}. Tangible objects, according to their definition, include people, locations, machines, and documents, while abstract objects encompass legal entities, organizational constructs, and electronic documents. 
At first glance, it may seem appropriate to align \texttt{OCED:Object} with \texttt{gufo:Object}, since the latter is defined as a concrete individual that persists through time while maintaining its identity \cite{Ontological_foundations_book}, and is characterized by existential independence \cite{Ontologically_Well_Founded_Profile}. However, the examples classified under \texttt{OCED:Object}, such as \textit{contract}, suggest a broader scope than that covered by \texttt{gufo:Object}. Therefore, we propose mapping \texttt{OCED:Object} to \texttt{gufo:Endurant}, a more general category that encompasses \texttt{gufo:Object} and includes all concrete individuals that endure through time, potentially undergo qualitative changes while preserving identity, and may depend existentially on other endurants, i.e., \texttt{aspects}, such as a contract depending on its signatories.

\vspace{0.2em}
\noindent$\blacktriangleright$ \texttt{Object Type}.
As stated in \cite{TowardsStandardOCED}, each object in the Core Model is associated with exactly one object type. Accordingly, we propose mapping \texttt{OCED:ObjectType} to \texttt{gufo:EndurantType}, which includes several subtypes such as \texttt{gufo:Kind}, \texttt{gufo:Phase}, and \texttt{gufo:Role}\cite{guizzardi2021types}. A \texttt{gufo:Kind} is a type that necessarily classifies its instances, providing a uniform identity for them (e.g., \textit{Person}). In contrast, \texttt{gufo:Phase} and \texttt{gufo:Role} are anti-rigid sortals—types that 
apply to their instances only under certain conditions and rely on an underlying kind to supply their identity criteria (e.g., \textit{Teenager}, \textit{Student}). This mapping from \texttt{OCED:ObjectType} to \texttt{gufo:EndurantType} 
introduces semantic flexibility, allowing object types to reflect different modeling needs. 

\vspace{0.2em}
\noindent$\blacktriangleright$ \texttt{Object Attribute}.
Each object in the Core Model is associated with an arbitrary number of attribute–value pairs, where attributes (e.g., \textit{skin color}, \textit{weight}) describe specific properties of the object and are assigned corresponding values (e.g., \textit{white}, \textit{68}) \cite{TowardsStandardOCED}. 
\texttt{gufo:Quality}, a subtype of \texttt{gufo:Endurant}, captures this notion. Qualities exist in time while maintaining their identity. For example, when we say that the temperature of the patient is rising, it is not 37$^{\circ}$C  that is rising but something that preserves its identity and that changes in time; however, unlike \texttt{gufo:Object}, they are existentially dependent on a single concrete individual \cite{Ontological_foundations_book,gufo_github}. They can also be measured within a specific value space (e.g., the color spindle). However, if object attribute values change over time, we employ a specific subclass of \texttt{gufo:Situation} known as \texttt{gufo:Quality Value Attribution Situation} or \texttt{gufo:QVAS} to model the state of a quality of a particular entity having that value for a given temporal duration. In other words, this allows one to capture the period during which a particular quality value is attributed, thus enabling the tracking of changes in attribute values over time \cite{gufo_github}. By connecting this notion to \texttt{gufo:Quality}, it becomes possible to monitor the evolution of attribute values throughout the life-cycle of an object. In the case of employing \texttt{gufo:QVAS}, its type and value can be determined by \texttt{gufo:concerns Quality Type} and \texttt{gufo:concerns Quality Value}, respectively. Additionally, this gUFO concept allows for the assignment of both start and end times to the situation. 
This directly covers the feature of timed attributes that differentiates OCEL V2 from the Core Model, as OCEL V2 natively supports it \cite{OCEL2}.

\vspace{0.2em}
\noindent$\blacktriangleright$ \texttt{O2O Relationship}.
One of the advantages of the object-centric approach is its capacity to represent relationships between objects. More expressive models, such as OCEL V2, even enable the qualification of these O2O relations. In our model, O2O relations are modeled as a subclass of \texttt{gufo:Endurant}, called \texttt{gufo:Relator}. They are individuals with their own identity that have the ontological capacity to mediate connections between endurants (\cite{Endurant_Types_in_Ontology}, \cite{gufo_github}). For instance, consider the relationship between two objects: a \textit{Customer} and a \textit{Product}. The relation \textit{Customer purchased Product} can be represented by the reified entity \textit{Purchase}, which is modeled as a \texttt{gufo:Relator}. 
Modeling O2O as \texttt{gufo:Relator} offers several advantages. First, it allows for the representation of attributes associated with the O2O relation itself, such as whether a \textit{Purchase} is \textit{pending} or \textit{finalized}. 
Second, it enables the tracking of changes to these relationships over time. Third, it facilitates the modeling of interactions between these relations and relevant events. 
Lastly, while most metamodels for OCED support only binary relations, modeling O2O as \texttt{gufo:Relator} allows us to represent relationships involving more than two objects, as well as anadyc relations, i.e., relations whose arities change at the instance level (e.g., the supervision relation can be irreducibly construed an anadyc relation: binary in some cases, ternary in others, quaternary in yet other cases).

\vspace{0.2em}
\noindent$\blacktriangleright$ \texttt{Event}. Fahland et al. \cite{TowardsStandardOCED} define an event as an observable, atomic phenomenon that occurs at a single point in time. However, later in the report, the authors acknowledge this atomic characterization as a limitation of current OCED metamodels. 
By contrast, the definition of events in gUFO addresses this limitation. A \texttt{gufo:Event} is a \texttt{gufo:ConcreteIndividual} that \textit{occurs} or \textit{happens} over time and involves or impacts \texttt{gufo:Endurants}. Events in this framework can be either instantaneous or extended over a temporal interval. The duration of an event is specified by its association with two points in time: the beginning and the end of the event \cite{gufo_github}. 
This ontological perspective enables events to be modeled as either atomic or composed of multiple (atomic or complex) sub-events \cite{ontology_of_events}. For example, in this model, the event of \textit{finalizing an order} may be broken down into atomic events such as \textit{reviewing items} and \textit{making a payment}. The classification of an event as atomic or composite depends on the modeling context and the specific analytical domain. For instance, \textit{making a payment} may itself be treated as a complex event in a different modeling scenario. The strength of the ontological approach lies in its capacity to support conceptual flexibility, allowing for both atomic and composite representations of events according to the requirements of the domain and metamodeling goals.

\vspace{0.2em}
\noindent$\blacktriangleright$ \texttt{Event Type}.
Every event is an instance of exactly one event type, which indicates the nature of the observation it represents. In most scenarios, the event type corresponds to the process activity that was executed 
\cite{TowardsStandardOCED}.
This concept corresponds to \texttt{gufo:EventType}, which is a \texttt{gufo:Type} whose instances are events (e.g., \textit{Earthquake}, \textit{Musical Performance}) 
\cite{gufo_github}.

\vspace{0.2em}
\noindent$\blacktriangleright$ \texttt{Event Attribute}.
Each event may possess an arbitrary number of attribute\-–value pairs
\cite{TowardsStandardOCED}. In our mapping, event attributes are represented using \texttt{gufo:\-Quality}, and their associated values are indicated via \texttt{gufo:hasQuality\-Value}. 
Unlike object attributes, event attributes are static and do not vary over time.

\vspace{0.2em}
\noindent$\blacktriangleright$ \texttt{E2O Relationship}.
The many-to-many relationship between events and objects is represented as \texttt{OCED:Observes} in the Core Model. 
According to this definition, in an object-centric context, an event does not merely observe an activity (i.e., event type), but also explicitly observes objects. In both the Core Model and OCEL V2, the association between an event and an object can be qualified, meaning the nature of their relationship is explicitly indicated. Common qualifiers include \texttt{CREATE}, \texttt{MODIFY}, and \texttt{DELETE}, although these are not mandated by the Core Model \cite{TowardsStandardOCED}. In gUFO, the relationships between endurants and events can be captured using \texttt{gufo:wasCreatedIn}, \texttt{gufo:wasTerminatedIn}, and \texttt{gufo:participatedIn} \cite{gufo_github}. 
The third relation, \texttt{gufo:participatedIn}, encompasses more than modification events. It also accounts for cases where objects are involved in an event without undergoing any change themselves \cite{Ontological_Considerations}. 

Given our mapping of static object attributes and O2O relationships to subclasses of \texttt{gufo:Endurant}, these relationships can also be established between such entities and events. This enables the modeling of events that influence static object attributes or relationships. For instance, a \textit{discount code} may be an attribute of an \textit{item}, which is terminated by an event such as \textit{ending a promotion}.  
Dynamic attributes are modeled using \texttt{gufo:QVAS} 
which can be linked to events via the property \texttt{gufo:broughtAbout}, indicating that the event brought about a specific attribute value. 
On the other hand, there are cases where an attribute value enables the occurrence of an event. For example, the event \textit{Buying} cannot occur if the value of the attribute \textit{Product availability} is zero. In these cases, the relation from \texttt{gufo:QVAS} to \texttt{gufo:event} should be modeled using \texttt{gufo:contributedToTrigger}.

\vspace{0.2em}
\noindent$\blacktriangleright$ \texttt{E2E Relationship}.
Structuring event data according to the OCED Core Model or OCEL v2 enables the inference of their order and \textit{directly-follows} relationships between events. However, 
events can be related in ways that go beyond their execution order. Yet mainstream business process modeling languages tend to focus almost exclusively on this one -admittedly essential- type of relationship \cite{Adamo2018BusinessPA}. 
We noted before that events may be atomic or complex. This implies a part–whole relationship between complex events and their atomic components, which can be tackled via the \texttt{gufo:isEventProperPartOf} object property \cite{gufo_github}. 
Furthermore each event is associated with two temporal values, a start and an end point, strictly ordered by a \texttt{precedes} relation. The range of possible temporal relations between two events corresponds to the interval relations proposed by Allen \cite{guizzardi2013towards}. Meaning, just by indicating the starting and ending points of events, gUFO derives the appropriate Allen relations between them. Another possible E2E relation is when one event triggers the start or end of another. For example, the event \textit{receiving payment} can initiate the event \textit{shipment handling}.
This type of E2E can be declared as historical dependence between events, using \texttt{gufo:historically Depends On}. Historical dependence is a relation between events that is stronger than mere accidental temporal ordering but weaker than strict causation (thus including the latter). Historical dependence is also transitive, hence the dependency can be traced beyond what is explicitly shown 
\cite{gufo_github}. 

Table~\ref{tab1} summarizes the main conceptual correspondences between OCED constructs and gUFO concepts.

\begin{table}[t]
     \centering
     \caption{Conceptual alignment between OCED and gUFO concepts}\label{tab1}
     \renewcommand{\arraystretch}{0.9}
   \resizebox{1.0\textwidth}{!}{%
     \begin{tabular}{|p{5cm}|p{9cm}|}
     \hline
     \textbf{Concept} & \textbf{Corresponding Concept in gUFO} \\
     \hline
     Object & Endurant\\
     Object Type & Endurant type\\
     Object Attribute & Quality\\
     Dynamic Object Attribute value & Quality Value Attribution Situation\\
     Object to Object Relation & Relator\\
     Event & Event, Atomic Event, Complex Event\\
     Event Type & Event Type\\
     Event Attribute & Quality\\
     Event to Object Relation & was Created In, was Terminated In, participated In\\
     Event to Event Relation & Event Proper Part Of, historical dependency, Allen relations \\
     \hline
     \end{tabular}
}
\end{table}

\section{Conceptual Assessment} 
\label{eval}

We first assess the practical relevance of our proposal through the characteristics framework introduced in~\cite{Goossens2024ObjectCentricEL}, developed for the evaluation of OCED models and log formats. With respect to the dimensions of ambiguity and expressiveness introduced at the beginning of the paper, we then examine how the proposed model addresses the challenges outlined in Section~\ref{OCEDchallenges} by means of two representative examples drawn from~\cite{TowardsStandardOCED}. The relevance of this explanatory test lies in the fact that these challenges are not artificially constructed, but rather generalisations of recurring real-world issues observed across several use cases and reported by the community~\cite{TowardsStandardOCED}. Moreover, these limitations have been acknowledged in recent works such as~\cite{Wei2025DirigoAM}, \cite{Khayatbashi2025AIEnhancedBP}, and \cite{Kretzschmann2025StateAwareOP}, each of which addresses one or more of them through targeted extensions tailored to specific problems. The emergence of such isolated extensions suggests that the identified challenges remain insufficiently resolved in current OCED metamodels.

\subsection{Evaluation of Structural Coverage}
\label{Evaluation_of_Structural_Coverage}
It is possible to highlight the distinctive features of our proposal by considering additional dimensions that are likewise empirical in nature and grounded in practical use cases. In what follows, we evaluate our model against the characteristics framework introduced in \cite{Goossens2024ObjectCentricEL}, which defines 19 criteria derived from three perspectives: process mining, object-centric process modeling, and database storage.  
Characteristics CH1–CH7 define fundamental properties that events and objects must satisfy (e.g., identifiers, types and attributes). These are explicitly supported in our metamodel. CH8 and CH9 concern traceable attribute changes and the existence of O2O, both of which are also supported.

The next three characteristics are particularly relevant, as OCEL V2, EKG, and the Core Model do not fully support them: CH10 (support for association classes between object relations), CH11 (support for dynamic object relations over time), and CH12 (support for object type inheritance). Regarding CH10, only OCEL V2 partially addresses this issue through relation qualifiers. In contrast, gOCED supports both CH10 and CH11 by employing \texttt{gufo:Relator} to model O2O relations as ontological entities that explicitly mediate (dynamic) connections between related objects. CH12 highlights the expressive power of a metamodel grounded in a foundational ontology. Although not explicitly instantiated in the current version of gOCED, object type inheritance is natively supported in gUFO and can be incorporated without semantic inconsistency.

Characteristics CH13–CH15 concern many-to-many relationships between events and objects; these are also supported in our model. The remaining four characteristics relate to data quality and storage aspects rather than conceptual modeling properties, and therefore fall outside the scope of this work.

\subsection{Addressing the Open Challenges}
The goal of this subsection is to show how gOCED can be used to address the issues outlined in~\autoref{OCEDchallenges} using two examples from~\cite{TowardsStandardOCED}, which illustrate limitations of existing models. The first example concerns a purchasing scenario, and we use it to discuss \textbf{C1}, \textbf{C2}, and \textbf{C3}. The second example is the one we introduced earlier in the paper, involving a student and two professors, and is used to discuss \textbf{C4}, \textbf{C5}, and \textbf{C6}. Throughout the discussion, we will highlight our comparison with respect to the dimensions of ambiguity and expressiveness introduced at the beginning of the paper.
 
\subsubsection{Addressing C1, C2, and C3}
\begin{figure}[t]
    \centering
    \includegraphics[width=0.75\linewidth]{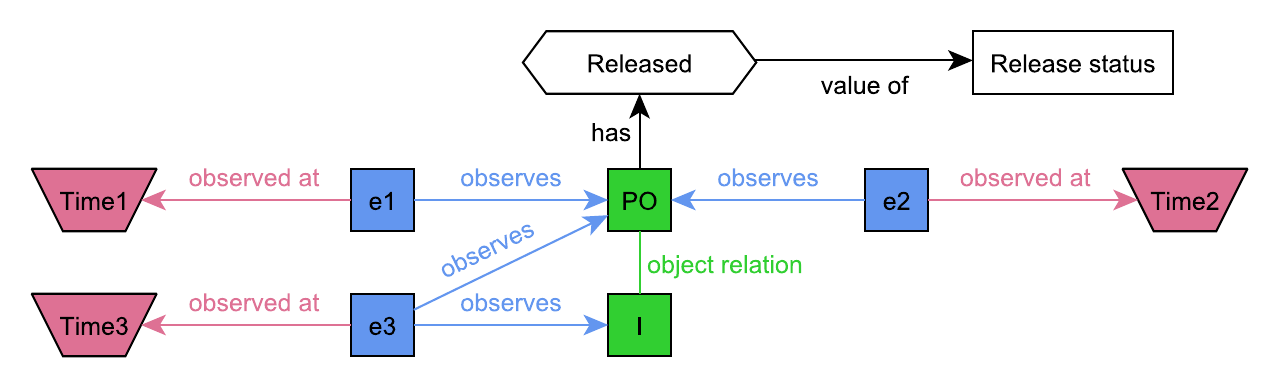}
    \caption{Example of ``Purchase order'' scenario modeled using the Core Model }
    \label{fig:purchaseScenario}
\end{figure}

The reference purchasing scenario is shown in~\autoref{fig:purchaseScenario} and involves an object, the \textit{Purchase Order (PO)}. Event \textit{e1} creates this object together with the attribute \textit{Release Status}, which its value is initially set to \textit{non-released}. The same \textit{PO} is then associated with another event, \textit{e2}, that releases it. This is reflected in the data as an update of the \textit{Release Status}'s value from \textit{non-released} to \textit{released}. Finally, \textit{e3} creates another object, the \textit{Invoice (I)}, in relation to the previously created \textit{PO}.

As previously mentioned, the Core Model does not provide a construct for modeling relationships between events. 
Some may argue that if the data of the above example were structured according to the Core Model, the directly-follows relation between \textit{e1} and \textit{e2} could be inferred, since both events are related to the same object (\textit{PO}), the timestamp of one precedes the other, and no other event related to the same object occurs between them \cite{Multiple_Behavioral_Dimensions_EKG}. Nevertheless, relationships between events cannot be reduced to only directly-follows relations. Also, it is important to recognize that the execution order of events may differ from their logical dependencies. Meaning execution order does not differentiate between an accidental temporal ordering from a necessary ontological (e.g., causal) dependence.


Consider the relationship between \textit{e2} and \textit{e3}. Depending on the vendor's behaviour, a receipt may only be issued after the order is released; alternatively, the order may be released only after a receipt is issued. In some cases, these events may not be causally related at all and may occur in any order. 
In addition, and this concerns in particular \textbf{C2}, the Core Model cannot represent complex events that are composed of other events.\footnote{This purchasing scenario does not include any complex event, but even in a simple example we may need to detect a complex event, such as a \textit{Successful Purchase}, which is composed of constitutive events such as \textit{releasing PO}, \textit{issuing invoice}, and \textit{paying}. In such a situation, the relation between \textit{Successful Purchase} and its sub-events would be modeled using \texttt{gufo:isEventProperPart}.}

The key point is that the Core Model and other current metamodels for OCED do not capture these distinctions, which effectively shifts the responsibility of inferring the correct process semantics to users when analyzing event logs. Assume that in this scenario a receipt will be issued only after the order is released. gOCED can capture, in an ontologically precise manner, the relationships between events and make this assumption explicit by connecting \textit{e2} and \textit{e3} through \texttt{gufo:historically Depends On}. Such constructs provide a standard way of modelling aspects that would otherwise require ad hoc solutions.


As demonstrated in~\cite{TowardsStandardOCED} the Core Model lacks support to represent object changes and to capture how different observations of the same object relate to each other. As a result, only the most recent value of the \textit{Release Status} attribute is displayed in~\autoref{fig:purchaseScenario}. This limitation can be addressed through three ad hoc approaches \cite{TowardsStandardOCED}: \begin{inparaenum}[\itshape (i)]
\item Event attributes that modify an object can be used to encode changes in its attribute values.
\item Object changes can be represented as a sequence of timestamped attribute values, where each value is directly linked to a specific time\footnote{This approach is employed by OCEL V2 \cite{OCEL2}}.
\item Using object snapshots, the same object \textit{o} can be observed multiple times, with each observation represented as a snapshot. The \texttt{OCED:observes} relation associates events with specific snapshots, and the same applies to the attributes and relations of \textit{o}.
\end{inparaenum}

\begin{figure}[t]
    \centering
    \includegraphics[width=1\textwidth]{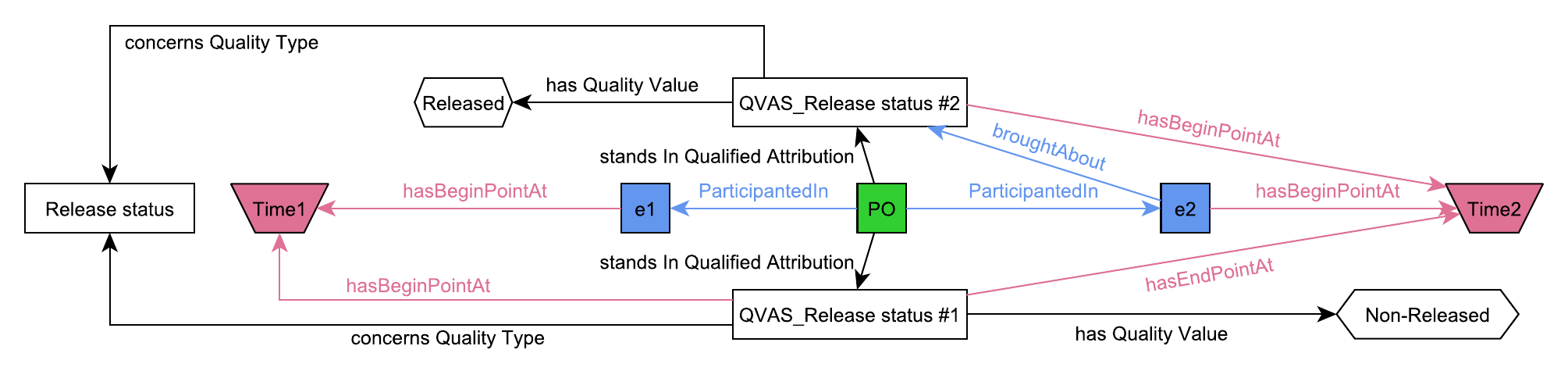}
    \caption{Attribute change modelled through gOCED
    }
    \label{fig:gufoObjectChange}
\end{figure}

These approaches 
result in several disadvantages. The first approach cannot capture value changes for multiple objects observed by a single event. With the second approach, determining the time interval during which a value \textit{v1} was valid requires assuming that its validity starts at time \textit{t1} and ends when the subsequent value of the same attribute, \textit{v2}, is observed at time \textit{t2}. Consequently, computing the validity interval of \textit{v1} relies on knowing \textit{t1}, the subsequent value \textit{v2}, and its timestamp \textit{t2}.
Finally, common sense rules out the third approach due to the large number of objects and the high storage it demands \cite{TowardsStandardOCED}.

gOCED, instead, offers the necessary constructs to represent dynamic object attributes. The changes in the attribute \textit{Release Status} in the Purchase Order scenario are shown in~\autoref{fig:gufoObjectChange}, which depicts both primary and secondary values. Another key aspect is 
that dynamic attributes, modeled as \texttt{gufo:QVAS}, can be directly connected to events. 
~\autoref{fig:gufoObjectChange} shows that the relation between \textit{e2} and the instances of \textit{Release Status} attribute can be modelled using the \texttt{gufo:broughtAbout}, which explicitly indicates that \textit{e2} changed the attribute value to \textit{Released}.\footnote{To obtain the same information in OCEL V2, we need to transitively re-link events and object attribute values through the timestamp of an event, its relation to an object instance, and the timestamps of that object’s attribute values~\cite{AbbRehse2024}.}

\texttt{gufo:QVAS} can be associated with beginning and end timestamps. This is beneficial in situations where the duration must be determined independently of the event that caused the change, or in cases where the attribute change is not triggered by any recorded event in the event log (e.g., a membership may expire simply due to the passage of time).

\subsubsection{Addressing C4, C5, and C6}
For the supervision scenario illustrated in~\autoref{fig:supervisionScenario}, gOCED models the O2O relationship between the student and the supervisors as an instance of \texttt{gufo:Relator}, denoted \textit{Supervision}. 

\begin{figure}[t]
     \centering
     \includegraphics[width=0.85\textwidth]{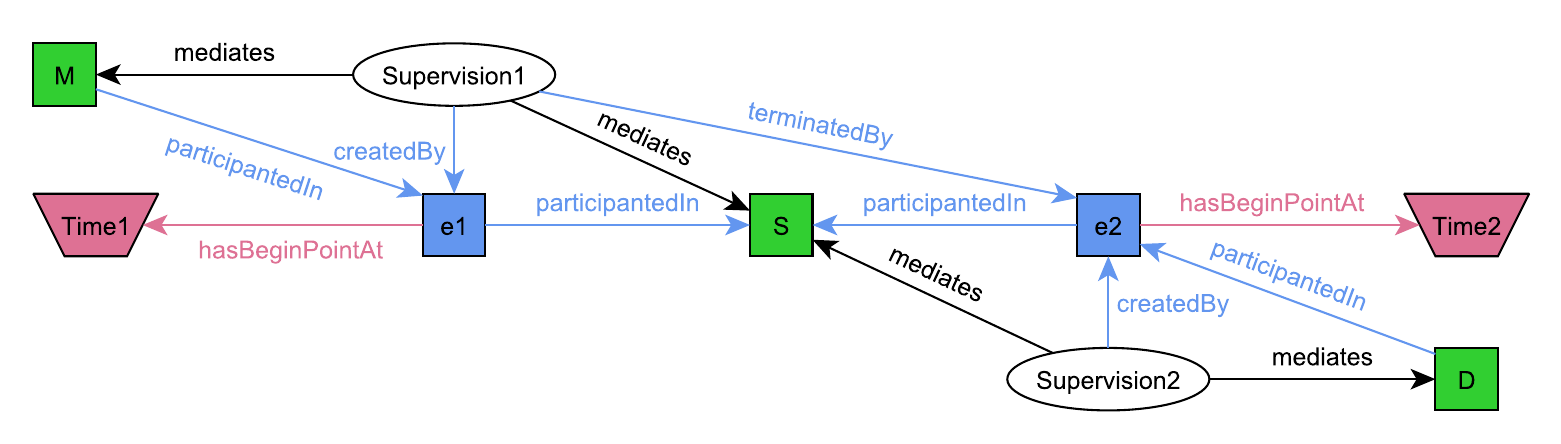}
     \caption{O2O modeling by gOCED - supervisor change}
     \label{fig:supervisionScenarioGUFO3.pdf}
\end{figure}

\begin{figure}[h!]
     \centering
     \includegraphics[width=0.9\textwidth]{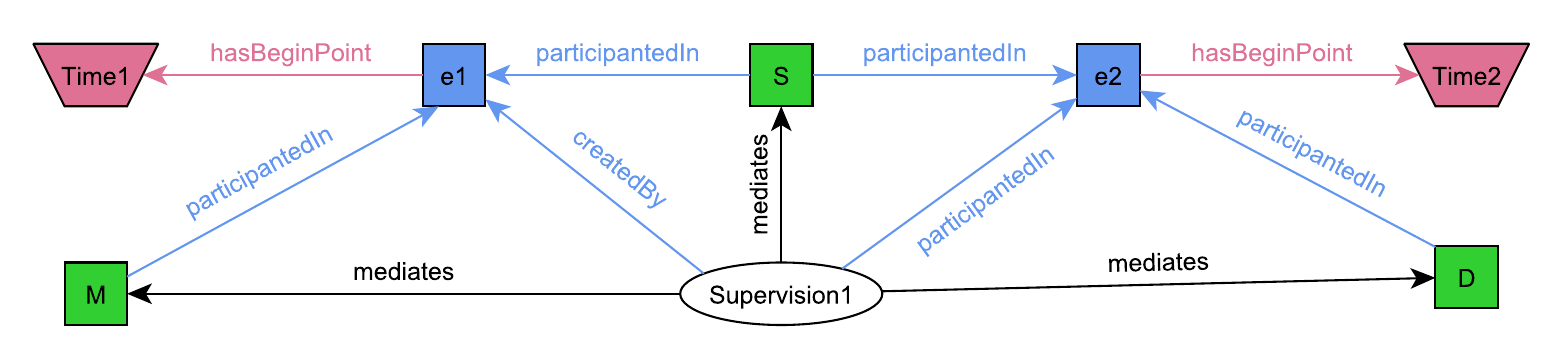}
     \caption{O2O modeling by gOCED - simultaneous supervision}
     \label{fig:supervisionScenarioGUFO2.pdf}
\end{figure}

Unlike the Core Model, gOCED yields a distinct structure for each circumstances discussed in~\autoref{introduction}. In the scenario shown in~\autoref{fig:supervisionScenarioGUFO3.pdf}, the supervisor changes from \textit{M} to \textit{D}, resulting in two separate \textit{Supervision} relators: one mediating between \textit{S} and \textit{M}, and another mediating between \textit{S} and \textit{D}. This explicitly captures the existence of two independent binary O2O relations. 
To model the occurrence of the second scenario (in which \textit{S} has two supervisors) a single \textit{Supervision} entity mediates between \textit{S}, \textit{M}, and \textit{D}, modeling an n-ary relation and indicating that both professors supervise \textit{S} simultaneously as shown in~\autoref{fig:supervisionScenarioGUFO2.pdf}.

\cite{TowardsStandardOCED} identifies another ambiguity in their model (also present in OCEL V2): when an event \textit{e} observes an object \textit{o1} with a relation \textit{R} to another object \textit{o2}, this only indicates that \textit{R} was observed at the time of \textit{e}, with no further interpretation possible. However, gOCED can represent the relationships between events and O2O explicitly. For example, as shown in both ~\autoref{fig:supervisionScenarioGUFO3.pdf} and ~\autoref{fig:supervisionScenarioGUFO2.pdf}, the link between \textit{e1} of type \textit{Assign Student} and an instance of \textit{Supervision} can be modelled directly, using \texttt{gufo:createdBy}, indicating that \textit{e1} has created that supervision.  Additionally, \texttt{gufo:Relator} makes it possible to have multiple distinct relations between same objects, as well as relators connecting relators (e.g., the mal-practice insurance connected to an employment; or cancellation insurance connected to an airline ticket).

gOCED can further enhance the expressiveness of the resulting event logs by allowing the inclusion of additional details. Consider an extension of the example in which the relationship between a student and her supervisor has a specific category, either thesis or internship. When \textit{Supervision} is modeled as a \texttt{gufo:Relator}, it becomes possible to associate a \textit{Category} attribute with this O2O relation or to extend a relator type with appropriate subtypes, thereby capturing this additional semantic distinction. 
gOCED also enables the explicit specification of time intervals for O2O relationships (and any other concrete individuals). Therefore, the start and end times of \textit{Supervision} instances can be directly defined, without needing to reconstruct these periods from a sequence of events.

\subsection{Simplicity}
So far, we have discussed the primary goal of our approach, namely its ability to extend contemporary models for object-centric event data towards handling the key challenges recalled in Section~\ref{OCEDchallenges}. We now briefly address simplicity, intended as follows: if one intends to represent an existing log, captured, e.g., as an OCEL or EKG, this can be done directly. In fact, all the concepts present in OCEL/EKG/OCED Core Model straightforwardly map into corresponding concepts within gOCED. 

Consider, for example an EKG $g$. Every object node in $g$ corresponds to a gOCED instance of \texttt{Endurant}. Every event node in $g$ maps to a gOCED instance of \texttt{Atomic Event} 
and gets, in fact, related to the corresponding timestamp attached to the event node in $g$. Every event-to-object relationship in $g$ becomes an instance of the \texttt{participatesIn} gOCED relationship. Every object-to-object relationship in $g$ becomes an instance of \texttt{Relator} in gOCED, for which no temporal property is specified - consistently with the fact that in EKG such relationships are atemporal. Similar considerations apply to OCELs (in view of the standard OCEL-to-EKG encoding \cite{Multiple_Behavioral_Dimensions_EKG}). 

\color{red}
\color{black}


\section{Empirical Evaluation 
} 
\label{empiricalEval}

The goal of this section is to assess whether the proposed model can represent OCED through a case study and whether the resulting representation can be constructed in a consistent and analyzable manner. To this end, the model was instantiated using the Northwind sample database\footnote{Northwind is a fictional Microsoft demo database representing a small trading company. In this study, we used a SQLite version derived from the original Microsoft sample database; see \url{https://learn.microsoft.com/en-us/power-apps/maker/canvas-apps/northwind-install} and \url{https://github.com/jpwhite3/northwind-SQLite3}.}, a relational dataset describing sales and order management processes. The dataset is object-centric by nature and includes interconnected business entities such as customers, employees, products, suppliers, shippers, territories, and regions, as well as transactional structures such as orders, order details, and employee-territory assignments. These characteristics make it a suitable case study for evaluating whether gOCED can capture necessary concepts such as objects, relators, events, qualities, and relations.

As a first step, we studied the dataset and mapped its concepts to those of the gOCED metamodel. To do so, we followed the definitions provided by gUFO. Concrete individuals that persist through time while maintaining their identity and are characterized by existential independence were mapped to \texttt{gOCED:object}, such as \emph{Product} and \emph{Customer}. In contrast, concrete individuals that also persist through time but are existentially dependent on other endurants and mediate among them, such as an \emph{Order}, were mapped to \texttt{gOCED:relator}. Occurrences that happen over time and involve or affect endurants were modeled as \texttt{gOCED:event}, such as \emph{OrderPlaced}. Among the attributes, those whose value changes were explicit in the source dataset, or whose temporal variability was considered relevant for the modeled domain, were treated as dynamic attributes and represented using \texttt{gOCED:QVAS}. Any source column not explicitly discussed in this mapping was treated as a static attribute and stored in the table of its corresponding owner. The identified concepts are summarized below:

\begin{itemize}
    \item \textbf{Objects:} Customer, Employee, Product, Supplier, Shipper, Territory, and Region.
    \item \textbf{Relators:} Order, LineItem, TerritoryAssignment, Supervision, Supply, and TerritoryInRegion.
    \item \textbf{Events:} OrderPlaced and OrderShipped.
    \item \textbf{Dynamic attributes:} Product UnitPrice, UnitsInStock, UnitsOnOrder, ReorderLevel, and Discontinued.
\end{itemize}

Based on this mapping, the transformed dataset was materialized into a set of schema-level and data-level tables. At the data level, each object type is represented by its own instance table, namely \texttt{customers\_data}, \texttt{employees\_data}, \texttt{products\_data}, \texttt{suppliers\_data}, \texttt{shippers\_data}, \texttt{territories\_data}, and \texttt{regions\_data}. Each of these tables contains an \texttt{object\_id} column used to identify the corresponding instance, together with its attributes. In a similar way, each relator type is represented in a dedicated table with a \texttt{relator\_id}, such as \texttt{orders\_data}, and\texttt{orderDetails\_data}. Events are stored in \texttt{event\_table}, where each event is identified by an \texttt{event\_id} and associated with its type and temporal information. Relations between events and endurants are stored in \texttt{event\_endurant\_data}, dependencies between events are stored in \texttt{event\_event\_data}, and dynamic attribute values are represented in \texttt{QVAS}.

The evaluation was conducted in three steps:
\begin{enumerate}
    \item Assessing the extent to which the source dataset could be represented in terms of objects, relators, events, and qualities, and reporting the size of the resulting instantiation.
    \item Checking the internal consistency of the generated tables and relations.
    \item Demonstrating the analytical usefulness of the resulting representation through example queries.
\end{enumerate}

\subsection{Coverage and Instantiation Statistics}

As a first step, we assessed both the coverage of the transformation and the size of the resulting instantiation. Coverage refers to the extent to which the source schema could be represented in terms of gOCED concepts, while instantiation statistics indicate the volume of the generated representation.

\begin{table}[t]
    \centering
    \caption{Coverage of the Northwind-to-gOCED transformation}
    \label{tab:coverage}
    \begin{tabular}{lr}
    \hline
    \textbf{Measure} & \textbf{Value} \\
    \hline
    Total source tables & 13 \\
    Mapped source tables & 11 \\
    Unmapped source tables & 2 \\
    Table coverage ratio & 0.846 \\
    Total columns in mapped tables & 84 \\
    Mapped columns & 82 \\
    Unmapped columns & 2 \\
    Column coverage ratio & 0.976 \\
    \hline
\end{tabular}
\end{table}

At the schema level, 11 out of the 13 source tables in the Northwind database were mapped to gOCED concepts. The mapped tables include \texttt{Customers}, \texttt{Employees}, \texttt{Products}, \texttt{Categories}, \texttt{Suppliers}, \texttt{Shippers}, \texttt{Territories}, \texttt{Regions}, \texttt{Orders}, \texttt{Order Details}, and \texttt{EmployeeTerritories}. In particular, \texttt{Categories} was not modeled as a separate object type; instead, its information was incorporated as static attributes of \texttt{Product}, namely category name, description, and picture. We did not map \texttt{CustomerDemographics} and \texttt{CustomerCustomerDemo} because these tables add very limited conceptual value to the target model. \texttt{CustomerDemographics} contains only an identifier and a textual description, while \texttt{CustomerCustomerDemo} functions merely as a bridge table between customers and demographic categories.

At the column level, 82 out of the 84 columns of the mapped source tables were explicitly represented in the transformation, corresponding to a column coverage ratio of 0.976. The two remaining unmapped columns can be explained by modeling scope decisions. In particular, \texttt{Employees.Photo} was excluded because it contains binary payload data that is not relevant for the intended analytical use, while \texttt{Categories.CategoryID} was not represented as a separate target element because the \texttt{Categories} table was incorporated into the model through static attributes of \textit{Product} such as category name, description, and picture.

\begin{table}[t]
    \centering
    \caption{Instantiation statistics of the Northwind-to-gOCED transformation}
    \label{tab:instantiation-stats}
    \begin{tabular}{lrr}
    \hline
    \textbf{Category} & \textbf{Types} & \textbf{Instances} \\
    \hline
    Objects & 7 & 268 \\
    Relators & 6 & 625{,}752 \\
    Events & 2 & 32{,}543 \\
    Dynamic attributes (QVAS) & 5 & 464 \\
    Event-endurant relations & -- & 81{,}368 \\
    Event-event relations & -- & 16{,}261 \\
    \hline
\end{tabular}
\end{table}


\subsection{Internal Consistency of the Generated Representation}

As a second step, we assessed the internal consistency of the generated gOCED instantiation. The goal of this step is to verify that the transformation preserves the structural dependencies implied by the source data and the mapping rules.

To this end, we performed a set of integrity checks over the generated tables. First, we verified that all references in \texttt{event\_endurant\_data} point to existing events in \texttt{event\_table}. Second, we checked that all source and target references in \texttt{event\_event\_data} correspond to existing events. Third, we verified that all entries in \texttt{QVAS} refer to valid owner instances. In addition, we checked the validity of the participants of each relator type, including orders, line items, territory assignments, supervision relations, supply relations, and territory-region relations. Finally, we performed a temporal sanity check to ensure that no \texttt{OrderShipped} event occurs before its corresponding \texttt{OrderPlaced} event.

All checks returned zero violations. This indicates that the generated representation is structurally coherent and that the transformation preserves the intended dependencies between objects, relators, events, and qualities. In particular, the absence of dangling references in event, relator, and QVAS tables suggests that the instantiated model is internally well-formed and faithful to the source data. Moreover, the temporal ordering constraint between \texttt{OrderPlaced} and \texttt{OrderShipped} was satisfied for all cases in which both events were present.

\subsection{Analytical Usefulness}

To illustrate the analytical usefulness of the instantiated metamodel, we formulated two example queries that rely on constructs that are made explicit in the proposed metamodel. The first query analyzes delayed fulfillment through the \textit{Order} relator. More specifically, it identifies delayed orders by comparing the time of \textit{OrderShipped} with the required date stored in the corresponding order, while also requiring the explicit historical dependency between \textit{OrderShipped} and \textit{OrderPlaced}. The results are then grouped by \textit{Customer}--\textit{Employee}--\textit{Shipper} relation which is an non-binary O2O. Meaning, this query does not treat these participants as isolated pairwise relations, but analyzes them jointly as roles mediated by the same relator.


The second query compares consecutive \texttt{UnitPrice} states of the same product and measures how demand changes from one state to the next. For each transition, it computes differences in the number of orders, total ordered quantity, and number of customers. This enables the analyst to observe how demand patterns vary across successive quality states and shows that the representation supports analyses over temporal change, not merely over current values or isolated events.

The second query compares consecutive \textit{UnitPrice} states of the same product and measures how demand changes from one state to the next. Its added value lies in the fact that the model represents \textit{UnitPrice} not as a static attribute, but as a temporally scoped quality captured through successive \textit{QVAS} states with explicit validity intervals. By aligning orders with the \textit{OrderPlaced} event time, the query can associate each transaction with the price state that was valid when it occurred. This makes it possible to analyze demand across successive quality states.

\section{Conclusion and Perspectives}
\label{conclusion}
This paper advances the research agenda toward establishing a standard metamodel for OCED, an essential foundation for object-centric process analysis and mining. Building upon the OCED Core Model, which integrates the main characteristics of various representation formats toward a unified standard \cite{TowardsStandardOCED}, our work addresses its key limitations and unresolved challenges. We do so by grounding the Core Model in a lightweight instantiation of the UFO-B ontology, namely gUFO. The resulting ontology-based model resolves several open challenges: it supports a correct treatment of binary O2O relations, provides a sound representation of events, enables the modeling of E2E relations, and captures changes to objects, attributes, and relationships over time. Furthermore, it permits the assignment of start and end times to all of these elements and offers explicit means to represent relationships between events and attribute/O2O changes.

An important aspect of this work's contribution is that, while providing a solution to the discussed open challenges, it also avoids the need for ad hoc extensions that address problems in isolation.
By grounding the metamodel in a foundational ontology, we establish a principled and coherent conceptual structure that ensures internal consistency while enabling greater expressiveness when needed. A clear example of this is the support for object type inheritance, as discussed in Section \ref{Evaluation_of_Structural_Coverage}. Although this feature is not explicitly instantiated in the current version of our metamodel, it is inherently supported by gUFO and can be incorporated without semantic conflict.

This work also contributes in practical terms, particularly with respect to event log extraction. Ontologies have previously been employed for extracting event data, for example through ontology-based data access approaches \cite{Calvanese2017OBDAFL}. Since process mining techniques require input data to be explicitly structured as event data \cite{Calvanese2017OBDAFL}, the quality and reliability of their results depend heavily on the structure and semantics of the underlying logs. Structuring event data according to an expressive and semantically grounded metamodel enables the unambiguous capture of all relevant information contained in the data, reducing information loss which is a key aspect to consider when manipulating OCED models \cite{TowardsStandardOCED}.

Finally, the model we propose allows for an alternative standard for the conceptualization of OCED. In contrast to existing approaches, which have largely evolved in a bottom-up manner, gOCED follows a top-down design grounded in a foundational ontology. This leads to a detailed yet flexible conceptual structure that can be systematically simplified (by removing undesired components) without compromising semantic integrity. A comparison with the Core Model, OCEL V2, and EKG shows that these models can be understood as reduced instantiations of our metamodel. This property enhances interoperability, as transformations between the models can be formally defined while preserving conceptual and semantic consistency. All in all, our approach enables a more precise and unambiguous representation of object-centric event logs in support of object-centric process mining. By doing so, we aimed at contributing to ongoing standardization efforts by complementing the current OCED metamodel with a comprehensive set of features that strike a balance between conceptual simplicity and expressive power.

This work is subject to some limitations. First, the proposed metamodel is grounded in gUFO/UFO-B and therefore adopts the ontological assumptions and distinctions of that framework. While this grounding provides semantic precision, it also means that alternative foundational choices may lead to different conceptualizations. Second, the contribution is intentionally focused on conceptual modeling aspects of OCED, in particular ambiguity reduction, expressiveness, and structural adequacy, rather than on storage, serialization, or data-quality concerns. Third, the validity of any instantiated model depends on the accuracy and completeness of the source data; while this assumption may hold in controlled evaluation settings, real-world data is often noisy, incomplete, or inconsistent, which may affect both model faithfulness and validation outcomes. Fourth, instantiation of the metamodel still requires modeling judgment, especially in borderline cases such as the distinction between objects and relators. Finally, the present work operates at the metamodel level; operational aspects such as tooling, automated transformations, and implementation support are outside the current scope.

These limitations also point to relevant directions for future work. First, we intend to exploit our approach towards devising novel process mining techniques that go beyond the assumptions made in the current ones (which, for example, assume that relationships do not change \cite{SWGM25}). In fact, the discovery of object-centric Petri net models capable of handling time-varying relationships (e.g., \cite{fahland2019describing,vanderwerfCorrectnessNotionsPetri2024,gianolaObjectCentricConformanceAlignments2024}) requires a proper treatment of dynamic O2O relations and, in turn, that the evolution of relationships is unambiguously represented in the event data, which our approach natively supports. 

Second, we aim to relax the current assumption that event data are accurate and complete. Addressing incomplete, uncertain, and ambiguous information is increasingly recognized as a critical challenge in process mining and remains largely unexplored in the object-centric context, with only a few preliminary contributions available \cite{FBSW25}.

\clearpage

\bibliographystyle{splncs04}
\bibliography{references}

\begin{thebibliography}{10}
\providecommand{\url}[1]{\texttt{#1}}
\providecommand{\urlprefix}{URL }
\providecommand{\doi}[1]{https://doi.org/#1}

\bibitem{xes}
{IEEE 1849-2016 XES Standard}. \url{https://xes-standard.org}

\bibitem{OCPM_Introduction}
van~der Aalst, W.M.P.: Object-centric process mining: An introduction. In: ICTAC Summer School, pp. 1--20. Publisher, Location (2021)

\bibitem{Unraveling_Fabric}
van~der Aalst, W.M.P.: Object-centric process mining: Unraveling the fabric of real processes. Mathematics  \textbf{11}(12), ~2691 (2023). \doi{10.3390/math11122691}

\bibitem{Dealing_with_divergence_convergence}
van~der Aalst, W.: Object-centric process mining: Dealing with divergence and convergence in event data. In: IEEE International Conference on Software Engineering and Formal Methods. pp.~1--2. IEEE (2019)

\bibitem{AbbRehse2024}
Abb, L., Rehse, J.R.: Process-related user interaction logs: State of the art, reference model, and object-centric implementation. Information Systems  \textbf{124},  102386 (2024). \doi{10.1016/j.is.2024.102386}

\bibitem{Adamo2018BusinessPA}
Adamo, G., Borgo, S., Francescomarino, C.D., Ghidini, C., Guarino, N., Sanfilippo, E.M.: Business process activity relationships: Is there anything beyond arrows? In: International Conference on Business Process Management (2018), \url{https://api.semanticscholar.org/CorpusID:52140810}

\bibitem{Framework_Extracting_Encoding}
Adams, J., Park, G., Levich, S., Schuster, D., van~der Aalst, W.: A framework for extracting and encoding features from object-centric event data. arXiv  \textbf{abs/2209.01219} (2022), \url{https://arxiv.org/abs/2209.01219}

\bibitem{allen1983maintaining}
Allen, J.F.: Maintaining knowledge about temporal intervals. Communications of the ACM  \textbf{26}(11),  832--843 (1983)

\bibitem{gufo_github}
Almeida, J.P.A., Guizzardi, G., Sales, T.P., Falbo, R.A.: gufo: A lightweight implementation of the unified foundational ontology (ufo). \url{http://purl.org/nemo/doc/gufo}, last accessed 2025/04/25

\bibitem{almeida2019events}
Almeida, J.P.A., Falbo, R.A., Guizzardi, G.: Events as entities in ontology-driven conceptual modeling. In: Conceptual Modeling: 38th International Conference, ER 2019, Salvador, Brazil, November 4--7, 2019, Proceedings 38. pp. 469--483. Springer (2019)

\bibitem{OC_PM}
Berti, A., van~der Aalst, W.: Oc-pm: Analyzing object-centric event logs and process models. International Journal on Software Tools for Technology Transfer  \textbf{25},  1--17 (2022)

\bibitem{OCEL2}
Berti, A., Koren, I., Adams, J.N., Park, G., Knopp, B., Graves, N., Rafiei, M., Liss, L., genannt Unterberg, L.T., Zhang, Y., Schwanen, T., Pegoraro, M., van~der Aalst, W.M.P.: {OCel (Object-Centric Event Log) 2.0 Specification} (2024), arXiv:2403.01975

\bibitem{botti2019representing}
Botti~Benevides, A., Bourguet, J.R., Guizzardi, G., Pe{\~n}aloza, R., Almeida, J.P.A.: Representing a reference foundational ontology of events in sroiq. Applied Ontology  \textbf{14}(3),  293--334 (2019)

\bibitem{Calvanese2017OBDAFL}
Calvanese, D., KALAYCI, T.E., Montali, M., Santoso, A.: Obda for log extraction in process mining. In: RW (2017), \url{https://api.semanticscholar.org/CorpusID:26448348}

\bibitem{TowardsStandardOCED}
Fahland, D., Montali, M., Lebherz, J., van~der Aalst, W., van Asseldonk, M., Blank, P., Bosmans, L., Brenscheidt, M., Di~Ciccio, C., Delgado, A., Calegari, D., Peeperkorn, J., Verbeek, E., Vugs, L., Wynn, M.T.: Towards a simple and extensible standard for object-centric event data (oced) - core model, design space, and lessons learned. arXiv  \textbf{abs/2410.14495} (2024)

\bibitem{Multiple_Behavioral_Dimensions_EKG}
Fahland, D.: Process Mining over Multiple Behavioral Dimensions with Event Knowledge Graphs, Lecture Notes in Business Information Processing, vol.~448. Springer, Cham (2022). \doi{10.1007/978-3-031-08848-3_9}

\bibitem{fahland2019describing}
Fahland, D.: Describing behavior of processes with many-to-many interactions. In: Application and Theory of Petri Nets and Concurrency: 40th International Conference, PETRI NETS 2019, Aachen, Germany, June 23--28, 2019, Proceedings 40. pp. 3--24. Springer (2019)

\bibitem{FBSW25}
Franceschetti, M., Buchegger, D., Seiger, R., Weber, B.: Toward iot-based process analytics: Extending event knowledge graphs with ambiguity. In: Proceedings of the 26th Edition of BPMDS (2025)

\bibitem{OCELV1}
Ghahfarokhi, A.F., Park, G., Berti, A., van~der Aalst, W.M.P.: Ocel: A standard for object-centric event logs. In: Bellatreche, L., et~al. (eds.) New Trends in Database and Information Systems. ADBIS 2021, Communications in Computer and Information Science, vol.~1450, pp. 169--175. Springer, Cham (2021). \doi{10.1007/978-3-030-85082-1_16}

\bibitem{ocelV1_web}
Ghahfarokhi, A., Park, G., Berti, A., van~der Aalst, W.: Ocel 1.0: Standard for object-centric event logs. \url{https://www.ocel-standard.org/1.0/specification.pdf} (January 2020), last accessed 2025/04/06

\bibitem{gianolaObjectCentricConformanceAlignments2024}
Gianola, A., Montali, M., Winkler, S.: Object-{{Centric Conformance Alignments}} with~{{Synchronization}}. In: Proc. 36th CAiSE. vol. 14663, pp. 3--19 (2024)

\bibitem{Goossens2024ObjectCentricEL}
Goossens, A., Smedt, J.D., Vanthienen, J.: Object-centric event logs: Characteristics, comparative analysis and road map. In: International Conference on Business Process Management (2024), \url{https://api.semanticscholar.org/CorpusID:273430888}

\bibitem{Goossens2023AligningOE}
Goossens, A., Verbruggen, C., Snoeck, M., Smedt, J.D., Vanthienen, J.: Aligning object-centric event logs with data-centric conceptual models. In: BPMDS/EMMSAD@CAiSE (2023), \url{https://api.semanticscholar.org/CorpusID:259167454}

\bibitem{DOCEL}
Goossens, A., De~Smedt, J., Vanthienen, J., van~der Aalst, W.M.P.: Enhancing data-awareness of object-centric event logs. In: Montali, M., Senderovich, A., Weidlich, M. (eds.) Process Mining Workshops. ICPM 2022, Lecture Notes in Business Information Processing, vol.~468, pp. 11--23. Springer, Cham (2023). \doi{10.1007/978-3-031-27815-0_2}

\bibitem{Ontological_foundations_book}
Guizzardi, G.: Ontological foundations for structural conceptual models. Tech. rep., Centre for Telematics and Information Technology (Telematica Instituut), Enschede (2005)

\bibitem{Endurant_Types_in_Ontology}
Guizzardi, G., Fonseca, C., Benevides, A., Almeida, J., Porello, D., Sales, T.: Endurant types in ontology-driven conceptual modeling: Towards ontouml 2.0. In: International Conference on Conceptual Modeling. pp. 1--13. Springer, Heidelberg (2018), \url{https://api.semanticscholar.org/CorpusID:52841941}

\bibitem{Ontologically_Well_Founded_Profile}
Guizzardi, G., Wagner, G., Guarino, N., van Sinderen, M.: An ontologically well-founded profile for uml conceptual models. In: International Conference on Advanced Information Systems Engineering. pp. 1--13. Springer, Heidelberg (2004), \url{https://api.semanticscholar.org/CorpusID:15334197}

\bibitem{guizzardi2022ufo}
Guizzardi, G., Botti~Benevides, A., Fonseca, C.M., Porello, D., Almeida, J.P.A., Prince~Sales, T.: Ufo: Unified foundational ontology. Applied ontology  \textbf{17}(1),  167--210 (2022)

\bibitem{guizzardi2021types}
Guizzardi, G., Fonseca, C.M., Almeida, J.P.A., Sales, T.P., Benevides, A.B., Porello, D.: Types and taxonomic structures in conceptual modeling: a novel ontological theory and engineering support. Data \& Knowledge Engineering  \textbf{134},  101891 (2021)

\bibitem{Ontological_Considerations}
Guizzardi, G., Guarino, N., Almeida, J., Almeida, A.: Ontological considerations about the representation of events and endurants in business models. In: Dubois, E., Pohl, K. (eds.) Advanced Information Systems Engineering (CAiSE 2016). Lecture Notes in Computer Science, vol.~9694, pp. 20--36. Springer, Cham (2016). \doi{10.1007/978-3-319-45348-4_2}

\bibitem{guizzardi2016ontological}
Guizzardi, G., Guarino, N., Almeida, J.P.A.: Ontological considerations about the representation of events and endurants in business models. In: Business Process Management: 14th International Conference, BPM 2016, Rio de Janeiro, Brazil, September 18-22, 2016. Proceedings 14. pp. 20--36. Springer (2016)

\bibitem{Guizzardi2008WhatsIA}
Guizzardi, G., Wagner, G.: What's in a relationship: An ontological analysis. In: International Conference on Conceptual Modeling (2008), \url{https://api.semanticscholar.org/CorpusID:3034252}

\bibitem{guizzardi2013towards}
Guizzardi, G., Wagner, G., de~Almeida~Falbo, R., Guizzardi, R.S., Almeida, J.P.A.: Towards ontological foundations for the conceptual modeling of events. In: Conceptual Modeling: 32th International Conference, ER 2013, Hong-Kong, China, November 11-13, 2013. Proceedings 32. pp. 327--341. Springer (2013)

\bibitem{ontology_of_events}
Guizzardi, G., Wagner, G., Falbo, R.A., Guizzardi, R., Almeida, J.P.A.: Towards ontological foundations for the conceptual modeling of events. In: International Conference on Conceptual Modeling. pp. 1--13. Springer, Heidelberg (2013), \url{https://api.semanticscholar.org/CorpusID:9223003}

\bibitem{Khayatbashi2025AIEnhancedBP}
Khayatbashi, S., Sj{\"o}lind, V., Gran{\aa}ker, A., Jalali, A.: Ai-enhanced business process automation: A case study in the insurance domain using object- centric process mining. ArXiv  \textbf{abs/2504.17295} (2025), \url{https://api.semanticscholar.org/CorpusID:278033577}

\bibitem{Kretzschmann2025StateAwareOP}
Kretzschmann, D., Berti, A., van~der Aalst, W.M.P.: State-aware object-centric process mining: Enhancing ocel 2.0 with explicit state transitions. In: IEEE International Enterprise Distributed Object Computing Conference (2025), \url{https://api.semanticscholar.org/CorpusID:285305395}

\bibitem{Uncovering_OCED}
Rebmann, A., Rehse, J.R., van~der Aa, H.: Uncovering object-centric data in classical event logs for the automated transformation from xes to ocel. In: Business Process Management. BPM 2022. Lecture Notes in Computer Science, vol. 13420, pp. 379--396. Springer, Cham (2022). \doi{10.1007/978-3-031-16103-2_25}

\bibitem{SWGM25}
Seidel, A., Winkler, S., Gianola, A., Montali, M., Weske, M.: To bind or not to bind? discovering object-centric processes under stable relationships. In: Proc. ER 2025 (2025)

\bibitem{Implementing_OCED_EKG}
Swevels, A., Fahland, D., Montali, M.: Implementing object-centric event data models in event knowledge graphs. In: De~Smedt, J., Soffer, P. (eds.) Process Mining Workshops. ICPM 2023. Lecture Notes in Business Information Processing, vol.~503, pp. 1--13. Springer, Cham (2024). \doi{10.1007/978-3-031-56107-8_33}

\bibitem{vanderwerfCorrectnessNotionsPetri2024}
{van der Werf}, J.M.E.M., Rivkin, A., Montali, M., Polyvyanyy, A.: Correctness {{Notions}} for {{Petri Nets}} with {{Identifiers}}. Fundamenta Informaticae  \textbf{190}(2-4),  159--207 (2024)

\bibitem{Verbruggen2024iDOCEMDA}
Verbruggen, C., Goossens, A., Smedt, J.D., Vanthienen, J., Snoeck, M.: idocem: defining a common terminology for object-centric event logging and data-centric process modelling. Softw. Syst. Model.  \textbf{24},  9--33 (2024), \url{https://api.semanticscholar.org/CorpusID:271104060}

\bibitem{Wei2025DirigoAM}
Wei, J., Ouyang, C., Wang, Y., Huang, L.: Dirigo: A method to extract event logs for object-centric processes. Data Knowl. Eng.  \textbf{160},  102485 (2025), \url{https://api.semanticscholar.org/CorpusID:279959162}

\end{thebibliography}
%

\end{document}